# Direct creation of micro-domains with positive and negative surface potential on hydroxyapatite coatings

T. Plecenik[a], S. A. M. Tofail[b], M. Gregor[a], M. Zahoran[a], M. Truchly[a], F. Laffir[b], T. Roch[a], P. Durina[a], M. Vargova[c], G. Plesch[c], P. Kus[a], A. Plecenik[a]

[a]Dept. of Experimental Physics, Faculty of Mathematics, Physics and Informatics, Comenius University, Bratislava, Slovakia
[b]Materials & Surface Science Institute, University of Limerick, Limerick, Ireland
[c]Department of Inorganic Chemistry, Faculty of Natural Sciences, Comenius University, Bratislava, Slovakia

A method for the direct patterning of electrostatic potential at the surface of hydroxyapatite is presented here. Micro-domains of surface potential have been created on hydroxyapatite coatings by a 20 keV focused electron beam with minimal alterations of surface chemistry. The success of such approach has been confirmed by Kelvin Probe Force Microscopy measurements, which show that this method is capable of creating micron sized positive and negative local electrostatic potential. The shape and potential difference of these domains were found to depend on the dose of total injected charge from the electron beam as well as the speed with which such charge is injected.

Modifications of biomaterials surfaces in terms of their wettability and surface potential (SP) play an important role in biological and medical applications as these are considered to be key factors that influence the adhesion and the growth of proteins, cells and bacteria on implant surfaces[1,2]. Various methods have been used to modify surface properties of biomaterials in order to control its interactions with proteins and biological cells. These include *inter alia* the application of an external electric field[2] (e.g. contact poling), ion or electron beam[3-7], photon irradiation[8], low and high pressure plasma[8], and chemical and electrochemical modifications. Of particular interest is the method proposed by Aronov *et al.* where such surface modifications have been obtained through a low-energy (100 eV) electron irradiation method[4-7]. These authors have demonstrated an ability to adjust SP and wettability (and thus also protein and bacterial adhesion) of hydroxyapatite (HA) in a wide range of contact angles. These authors irradiated either the whole sample surface or large areas (500 μm diameter) of HA ceramic pellets through a pre-patterned Si-mask and reported the creation of a negative SP only.

Here we report a convenient technique of direct microscopic patterning of SP i.e. without the need of any mask. In this technique, micro-domains of modified SP have been created on hydroxyapatite ($Ca_{10}(PO_4)_6(OH)_2$) coatings by focused 20 keV electron beam, typically available as an electron microprobe in a Scanning Electron Microscope (SEM). In contrast with the earlier investigations[4-7], our work shows that by varying the dosage and speed of electron injection it is possible to create circular, doughnut shaped, and bimodal micrometer-sized patterns with positive and negative SP.

Hydroxyapatite (HA) thin films have been deposited on Si substrates by spin-coating through a sol-gel synthesis route. For the sol-gel preparation, phosphoric pentaoxide ($P_2O_5$) and calcium nitrate tetrahydrate ($Ca(NO_3)_2 \cdot 4H_2O$) have been dissolved in pure ethanol to prepare a 0.5 mol/l and a 1.67 mol/l solutions, respectively. The precursor sol is then obtained by mixing the two solutions in a proportion to obtain a Ca/P molar ratio of 1.67. The mixture is continuously stirred at room temperature and subsequently heated to 70 °C for 1 hour, resulting in a transparent sol-gel, which is spin coated on a p-type (100) Si at 4000 rpm for 50 seconds. The deposited film on Si is then aged at 100 °C temperature for 30 minutes and calcined in air at 700 °C for 1 hour. The phase purity and the surface concentration of the calcined film have been verified by X-ray diffraction (XRD: Philips X'Pert Pro) and X-ray photoelectron spectroscopy (XPS: Kratos Axis 165).

The thickness of the films has been determined in cross-section with a Scanning Electron Microscope (SEM) TESCAN TS 5136 MM to be about 650 nm. The film surface morphology has also been inspected with SEM. For the electron beam modification, a matrix of 2x2 points has been irradiated by focused electron beam with an absorbed beam current ranging from 1.4 to 14 nA and a beam diameter ranging from 120 nm for the lowest current (1.4 nA) to 750 nm for the highest current (14 nA) used. The beam exposure time has also been varied from 1 to 700 seconds/exposure point. In all cases, the electron energy is kept constant at 20 keV at a vacuum of $10^{-4}$ Pa.

A Scanning Probe Microscope (SPM: NT-MDT Solver P47 PRO) in a semi-contact Atomic Force Microscopy (AFM) and Kelvin Probe Force Microscopy (KPFM) modes has been used for the measurements of the surface topography and SP of the HA films, respectively. TiN-coated silicon AFM probes have been used for KPFM measurements, which simultaneously provided topographic images. The AFM surface topography images showed the nanocrystalline nature of the calcined film although some nanoscale porosity was also visible. This nanocrystalline morphology has further been confirmed by SEM and XRD.

Table I compares the surface chemistry of the untreated and irradiated HA films. In the latter case, an array of 80x80 micro-domains was directly created over an area of 1 mm². Each of these micro-domains was created by exposing to a 20 keV focused electron beam with a 14 nA beam current. The exposure time was varied for 1 second and 7 seconds. The change in surface chemistry due to the electron irradiation is negligible and is mainly confined within the adsorbed over layer.

The SP distribution of these patterned areas has been measured by KPFM and some typical SP patterns are shown in Fig. 1. No topographic changes due to the electron beam exposure were observed by AFM (Fig. 2a,b) or SEM. Most importantly, all beam currents and exposure times used in the treatment have resulted in the creation of micro-domains with modified SP albeit with different types of SP distribution. Typical SP difference varied approximately from 150 to 400 mV.

We found that the diameter of these domains depended on the total dosage of the applied charge, Q. The lower dosage (Q ≈ 9,8-98 nC) resulted in patterns with an outer diameter of the SP of about 5-7 µm (Fig. 1a, d and e), while for the higher dosage (Q ≈ 98-980 nC) the outer diameter of the SP domains increased to about 8-10 µm (Fig. 1b, c and f). This is significantly higher than the minimal size of domains given by interaction area of about 3 µm diameter estimated by us using Monte Carlo simulations.

Similar spreading effects have been observed by Molina et al.[9] and He et al.[10] who used focused electron beam to create ferroelectric domain structures. Molina et al.[9] have attributed such effect to a migration of charge carriers inside the material caused by local electric fields induced by the irradiation. On the other hand, the electron beam can be significantly de-focused when the sample surface is charged. This could be an additional factor responsible for broadening of the domains at higher exposure times. There is also a possibility of the progressive spread of thermal damage although XPS results in Table I rules out any such effects.

Interestingly, the speed of charge injection, measured as the absorbed beam current greatly influences the resulting shape and distribution of the SP. The lower absorbed currents (1.4 nA) and thus slower charge injection are capable of creating domains only with negative SP (Fig. 1a – 1c) with respect to the unexposed surroundings, the SP of which is referenced as a zero potential. A similar amount of charge, when injected faster (14 nA absorbed beam current), creates domains with a bimodal distribution of positive (Fig. 1d – 1e) or less-negative (Fig. 1f) SP surrounded by a negative annulus of SP. Aronov et al.[6] have also observed negative SP in HA but in areas exposed to low energy (100 eV) electron irradiation. They propose that the impacting electrons generate electron-hole pairs, which are subsequently localized (together with primary electrons) by traps of different origin. As electrons have higher mobility, they may occupy bulk traps, or in our case, further surface traps. Holes can be trapped only by surface traps within close vicinity of the irradiation region, which in our case is hundreds of nanometers in diameter. This may lead to the bimodal SP distribution observed in Fig. 1d-e. Moreover, the emission of secondary electrons may leave positively charged holes at the surface[11] and give rise to positive SPs as observed in Fig 1d-e when HA was irradiated with electron beam with relatively higher charge injection. The intensity of these positively charged areas increased initially (Fig 1e). If the domains were exposed for sufficiently long time, incoming electrons could neutralize these positively charged areas although leaving them still less negatively charged than the areas where no such positive charge was created (Fig. 1f). Furthermore, local electric field-induced polarization[12] and creation of carbon/hydrocarbon surface contamination layer in SEM[4] may influence the resulting SP distribution. While more fundamental study will be necessary to distinguish the influence of a particular process in creating these domains, the creation of positively charged areas will be interesting to study specific binding of negatively charged proteins and cell membranes.

*Table I: XPS surface chemical analysis of HA thin films untreated and irradiated by electron beam (20 keV, 14 nA) for 1 and 7 seconds.*

| Elements | bonding type | Untreated (%) | E beam treated (%) 1 sec. | E beam treated (%) 7 sec. |
|---|---|---|---|---|
| C | C-C | 17.8 | 22.6 | 25.7 |
|   | C-O | 6.5 | 1.5 | 1.7 |
|   | O-C=O | 3.5 | 1.1 | 1.3 |
| O | O in HA | 34.9 | 37.1 | 36.2 |
|   | O=C | 6.7 | 5.5 | 4.6 |
|   | O-C | 4.3 | 2.9 | 2.3 |
| Ca |   | 15.9 | 17.8 | 17.0 |
| P |   | 10.4 | 11.6 | 11.3 |
| Ca/P ratio |   | 1.53 | 1.53 | 1.50 |
| O/Ca ratio in HA |   | 2.2 | 2.1 | 2.1 |

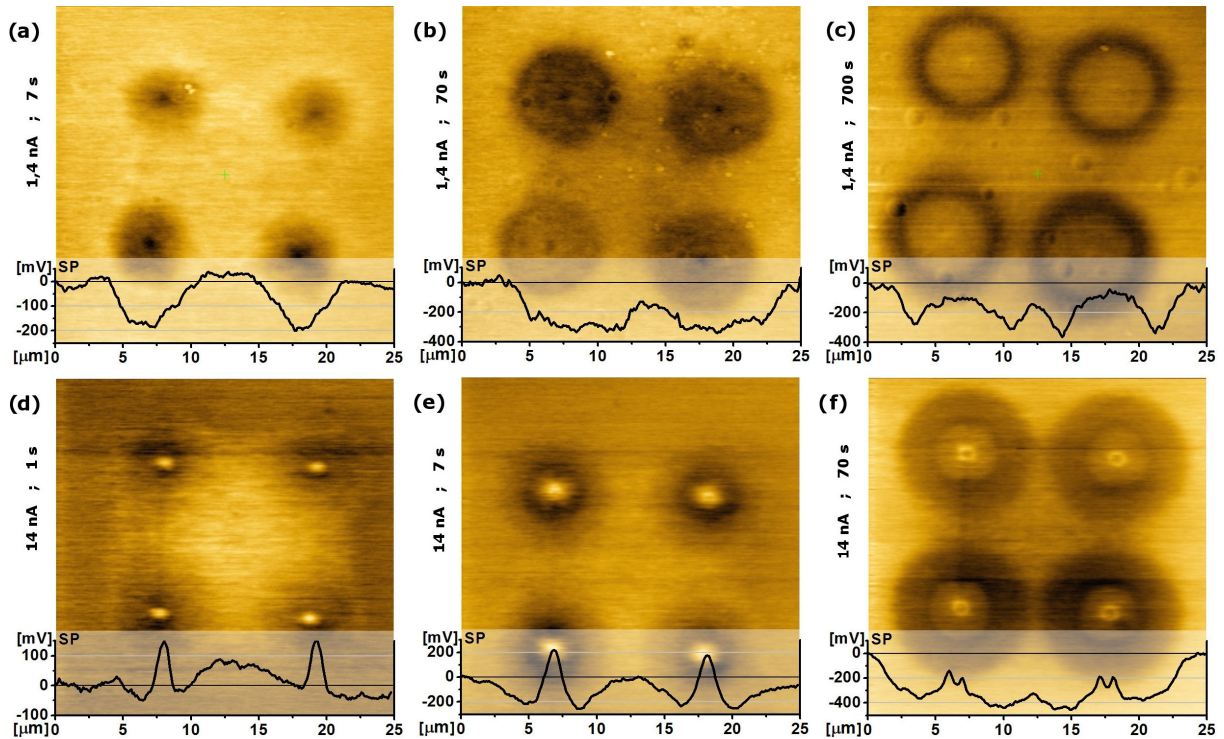

*Fig. 1: Surface potential distribution of areas irradiated by electron beam. Matrix of 2x2 points was irradiated in each area. Electron beam current and irradiation times for each point are show on the left side of every image. Surface potential profile of cross-section through 2 irradiated points is shown under each image The applied charge dosage: a) 9,8 nC; b) 98 nC; c) 980 nC; d) 14 nC; e) 98 nC and f) 980 nC*

These domains are stable in air when kept at room temperature. After one month, there was no significant difference in the size, shape or SP when compared to the measurements done immediately after the treatment (Fig. 2c). Thus, we expect that such domains can be stable for several months when stored in air and will facilitate practical applications. Such long-term stability is in agreement with the observations by Aronov et al.[4-7], where wettability and SP modifications of HA obtained by low-energy electron irradiation were stable for at least one month without any chemical or structural damage. Our study confirms these findings and extends to interactions with higher energy electron beam (20keV). It is known that structural alterations of HA can occur if exposed to 100-400 keV electron beam energy, e.g. as reported by Huaxia *et al.* during transmission electron microscopy[13]. XPS data presented in Table I establishes that such damage is unlikely to take place for the beam energy used in this study.

To conclude, micro-domains (5 – 10 μm diameter) with modified surface potentials were patterned on HA coatings by a focused 20 keV electron beam from a SEM microscope. We have shown that this method can conveniently create micro-domains with both positively and negatively modified surface potentials with respect to the untreated area. The size of these domains depends mostly on total electric charge injected, while the polarity and the shape depend on the speed of the charge injection (i.e. beam current). We have also shown that when stored in air, these domains are stable for at least one month.

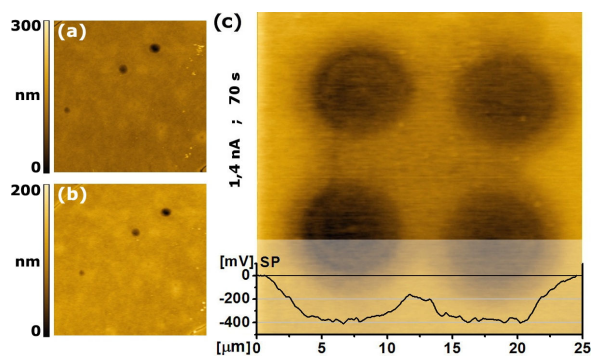

*Fig. 2: 25x25μm AFM topography of HA surface a) before electron beam irradiation b) the same area after irradiation with same parameters as in Fig. 1f. c) SP distribution after one month in the sample shown in Fig 1b.*

This project has been funded with support from the European Commission (EC NMP4-SL-2008-212533 – BioElectricSurface). This publication reflects the views only of the authors, and the Commission cannot be held responsible for any use which may be made of the information contained therein. This work was also supported by the Slovak Research and Development Agency under the contract No. DO7RP-007-09 and is also the result of the project implementation: 26240120012 supported by the Research & Development Operational Programme funded by the ERDF.